\documentclass[jkps,twoside,twocolumn]{revtex4} 
\usepackage{graphicx}
\begin{document}
\title[Waiting time]{Waiting-time distribution for a stock-market index} 
\author{Jae Woo  \surname{Lee},  }
\email{jaewlee@inha.ac.kr} 
\affiliation{Department of Physics, Inha University, Incheon, 402-751, Korea}
\affiliation{School of Computational Science, Florida State University, Tallahassee, FL 32306-4120, USA}
\author{Kyoung Eun \surname{Lee}, }
\affiliation{Department of Physics, Inha University, Incheon, 402-751, Korea}
\author{Per Arne \surname{Rikvold}} 
\affiliation{School of Computational Science, Department of Physics, Center for Materials Research
and Technology, and National High Magnetic Field Laboratory, Florida State University, Tallahassee, FL 32306-4120, USA}
\received{August 31 2005}

\begin{abstract}
We investigate the waiting-time distribution of the absolute return in the Korean stock-market
index KOSPI. We define the waiting time as a time interval during which the normalized absolute
return remains continuously below a threshold $r_c$. 
Through an exponential bin plot, 
we observe that the waiting-time distribution shows power-law behavior,  
$p_f (t) \sim t^{-\beta}$, for a range of threshold values. The waiting-time distribution 
has two scaling regimes, separated by the crossover time $t_c \approx 200$~min. 
The power-law exponents of the waiting-time distribution
decrease  when the return time $\Delta t$ increases. 
In the late-time regime, $t > t_c$, the power-law exponents are
independent of the threshold to within the error bars for fixed return time.
\end{abstract}


\maketitle

\newcommand{\be}{\begin{equation}}
\newcommand{\ee}{\end{equation}}

\section{INTRODUCTION}
In recent decades, the dynamics of stock markets have been studied by a number of  
methods from statistical 
physics \cite{MA99,MA97,BP00,SO03,MS95,BS94,BA1900,MA63,FA63,GP99,GM99,LG99,GP00,SA00,BC03,LE02,LL04,LL05}.
The complex behaviors of economic systems have been found to be very 
similar to those of other complex systems, customarily studied in statistical 
physics; in particular, critical phenomena. Stock-market
indexes around the world have been precisely recorded for
many years and therefore represent a rich source of data for quantitative
analysis. The dynamic behaviors of stock markets have been studied by
various methods, such as distribution functions \cite{GP99,GM99,LG99,GP00,LL04}, correlation
functions \cite{LG99,GP00,SA00},
multifractal analysis \cite{GD02,SK00,AR02,EK04,BE03,AI02,IA99,TP03,BE01,MA03,XG03,JWLEE05},
network analysis \cite{BC03},
and waiting-time distributions or first-return-time
distributions \cite{PM96,AL04,LA04,WL02,SW04,CO04,YD04,DP05,BE05,LJ05,IT95,SG00,RS00}.
 
Waiting-time distributions (wtd's) have been studied for many
physical phenomena and systems, such as self-organized criticality \cite{PM96}, rice piles \cite{AL04},
sand piles \cite{LA04}, solar flares \cite{WL02},
and earthquakes \cite{SW04,CO04,YD04,DP05,BE05,LJ05,IT95}.
Studies of wtd's have also been performed for
many high-frequency financial data sets \cite{SG00,RS00,RS02,SG03,SK02,KK04}.
Concepts of the continuous-time random walk (CTRW) have also been applied to 
stock markets \cite{SG00,RS00,RS02,SG03,SK02}. 
A power-law distribution for the calm time intervals of the price changes has been observed
in the Japanese stock market \cite{KK04}.

In the present work we consider the wtd for the  
Korean stock-market index KOSPI (Korean Composite Stock Price Index).
A waiting time of the absolute return
is defined as an interval between a time when the absolute return falls below 
a fixed threshold $r_c$, and the next time it again exceeds $r_c$. It therefore corresponds 
to a relatively calm period in the time series of the stock index. 
We observed power-law behavior of the 
wtd over one to two decades in time. 

The rest of this paper is organized as follows. 
In Section II, we introduce the return of the stock index and its
probability density function. In Section III, we present the wtd.
Concluding remarks are presented in section IV.

\section{Return of the Stock Index}
We investigate the returns (or price changes) of the Korean stock-market
index KOSPI.
The data are recorded every minute of trading from March 30, 1992,
through November 30, 1999 in the Korean stock market. 
We count the time during trading hours and remove closing hours, weekends, and
holidays from the data. Denoting the stock-market index as $p(t)$, the logarithmic
return is defined by
\be
   g(t)= \log p(t) - \log p(t-\Delta t) \;,
\ee
where $\Delta t$ is the time interval between two data points, the  
so-called return time. The logarithmic return $g(t)$ is thus a function of both $t$ and 
$\Delta t$. In this article we consider 
the return times $\Delta t=$~1min, 10~min, 30~min, 60~min, 600~min (=1day),
and 1200~min.
The normalized absolute return is defined by
\be 
  r(t) = \left \vert \frac{g(t)-\langle g(t) \rangle }{\sigma(\Delta t)} \right \vert
  \;,
\ee
where $\sigma(\Delta t)$ is the  standard deviation and $\langle \cdots \rangle$ denotes
averaging over the entire time series.
It is well known that the probability distribution function (pdf)  of the return 
$g(t)$ has a fat tail \cite{GP99,GM99}.
The tail of the pdf obeys a power law,
\be
p(x) \sim x^{-(1+\alpha)} \;,
\ee
where $\alpha$ is a nonuniversal scaling exponent that depends on the return
time $\Delta t$. The cumulative pdf then also follows a power law, such that
\be
P(g>x)= \int^{\infty}_{x} p(y) dy \sim x^{-\alpha} \;.
\ee
We observed clear power-law behavior in the tail of 
the pdf. Using least-squares fits, we obtained the power-law exponents $\alpha=3.06(8)$ for $\Delta t=1$~min and
$\alpha=3.2(4)$ for $\Delta t=600$~min.

\begin{figure}[t!]
\includegraphics[width=7cm,angle=0,clip]{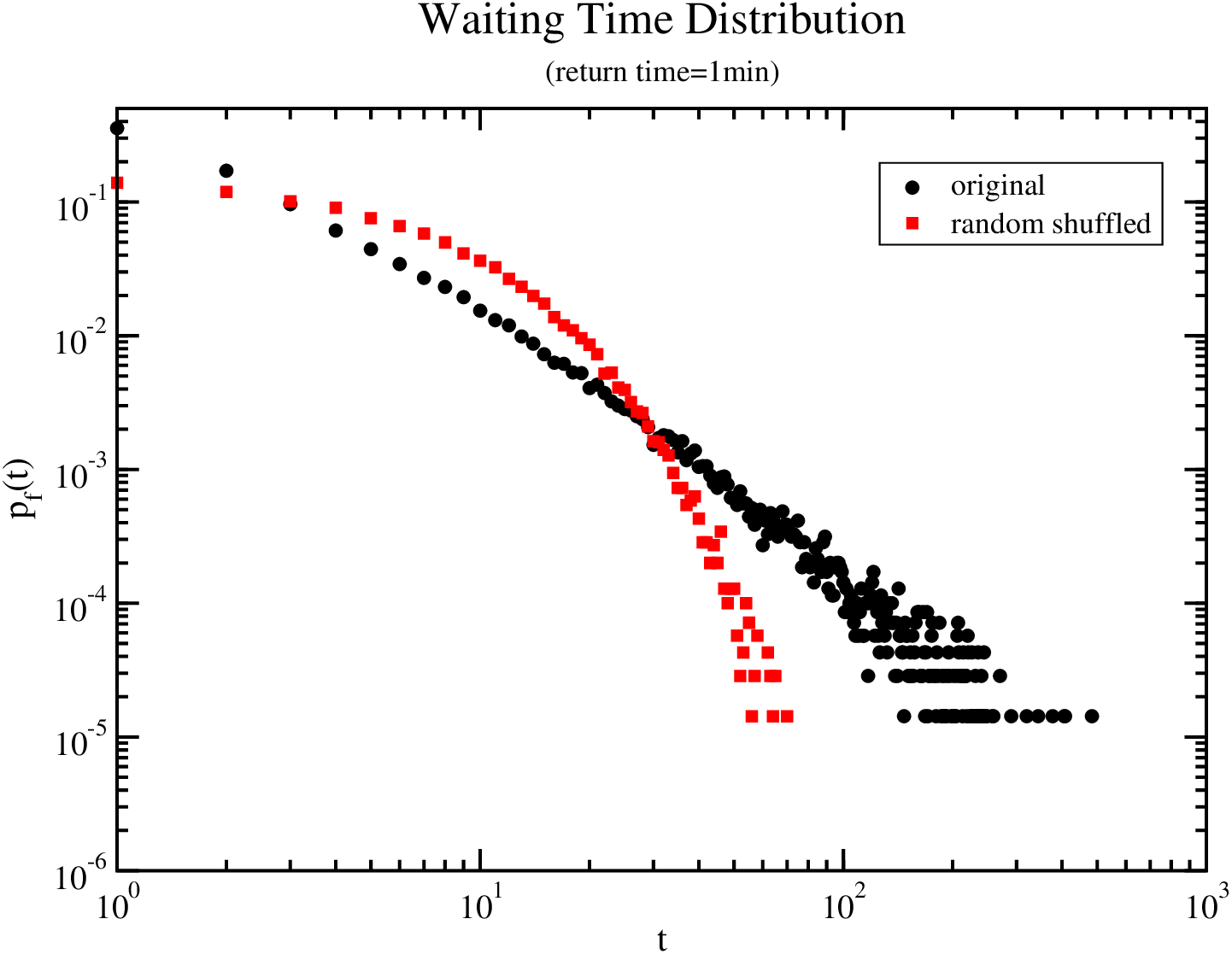} 
\caption[0]{
(Color online.)
Waiting-time distribution (wtd) of the absolute return for the original 
data (circles)  and the randomly shuffled data (squares) with $\Delta t=1$~min and $r_c=1$.
}
\label{fig1}
\end{figure}

\section{Waiting-time Distribution}

Consider a time series of the absolute return in the stock-market index. 
The waiting time of the absolute return with a threshold $r_c$ 
is defined as an interval between a time when the absolute return falls below 
a fixed threshold $r_c$, and the next time it again exceeds $r_c$. 
It  corresponds to a  calm period in the time series of the stock index. 
The waiting times depend on the threshold $r_c$ and the pdf of the absolute 
return. 
For small return times, for example $\Delta t=1$~min as in Fig. \ref{fig1}, 
the absolute return is distributed in a wide range
up to $r=500$. However, for large return times, the absolute return is
distributed in a narrow range. 
For large values of the threshold $r_c$, the waiting time has very
long time intervals. For small values of the threshold $r_c$, the
waiting time has many short time intervals. 

In Fig.~\ref{fig1} we present the wtd of the absolute return
for the original data set of KOSPI, together with a randomly shuffled data set. 
Both sets were analyzed with the threshold $r_c =1.0$ for the return time $\Delta t=1$~min. 
The randomly shuffled data were obtained by exchanging two randomly selected
return, repeating the exchanges one hundred times the
total number of data points.
The wtd of the absolute return shows the power law,
\be 
p_f (t) \sim t^{-\beta} \;,
\ee
where the scaling exponent $\beta$ depends on the return
time $\Delta t$.
However, the randomly shuffled
data lose the correlations of the original time series, 
and the uncorrelated wtd is therefore a simple 
exponential distribution, $p_f (t) = \frac{1}{\langle T \rangle } \exp(-t/ \langle T \rangle)$,
where $\langle T \rangle $ is the mean waiting time 
for the given threshold $r_c$. 

\begin{figure}[h!]
\includegraphics[width=7cm,angle=0,clip]{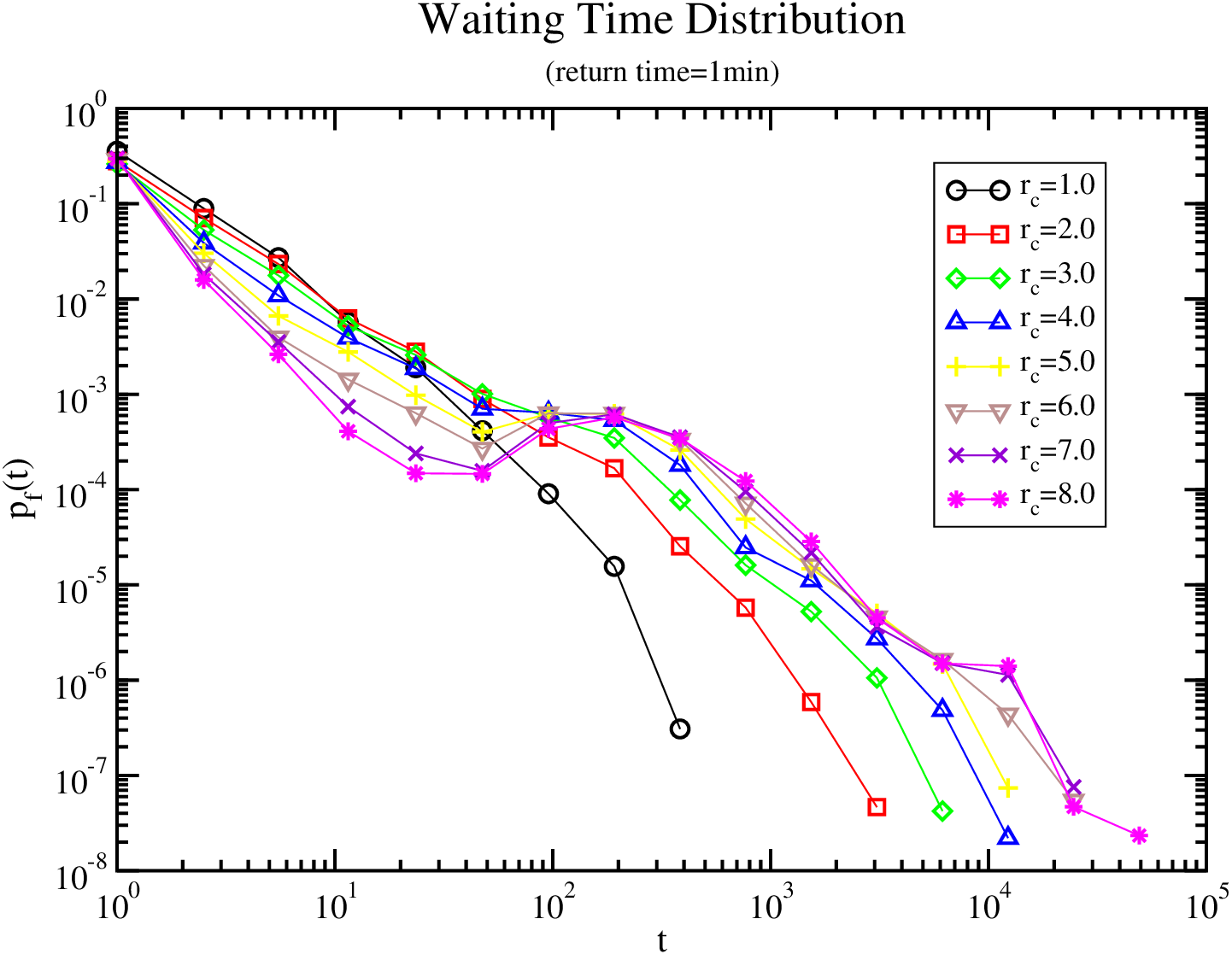}
\caption[0]{(Color online.) 
Wtd of the absolute return with $\Delta t=1$~min 
and several values of $r_c$, using the exponential bin plot.
}
\label{fig2}
\end{figure}

In the wtd in Fig.~\ref{fig1}, the data are sparsely distributed in the tail,
so it it difficult to measure the scaling exponents of the power law. 
To calculate the scaling exponents, we have therefore applied the exponential bin 
plot \cite{RZ05}.
In the exponential bin plot, we calculate the normalized histogram in 
bins of exponentially increasing size. If the distribution follows
a power law with exponent $\beta$, then the histogram of the distribution also has 
the same slope $\beta$ in the log-log 
exponential bin plot: $\log p_f (t) = C -\beta \log t$, where
$C$ is a constant depending on the return time and the threshold.

In Fig.~\ref{fig2} 
we present the wtd obtained using the exponential bin plot for the return time
$\Delta t=1$~min. 
We observe clear power-law behavior  
$t > t_c$, where $t_c$ is a crossover time. 
For a small return time, $\Delta t=1$~min, we observe 
two scaling regimes separated by a crossover time 
$t_c \approx  200$~min. In the log-log plots the
curves for $r_c =1.0$ to  $r_c =3.0$ are parallel to each other for
$t < t_c$, and the slope is measured as $\beta_1 =1.48(3)$. 
For $t< t_c$ and $r_c \ge 4.0$, the wtd decreases quicker than a power law
and shows a local maximum around $t=t_c$. 
When we choose a large threshold value $r_c$, for example $r_c =8.0$,
the wtd is still large for small waiting times. This
means that the return has clustering behavior, i.e., large
absolute returns occur in bursts.
For $t>t_c$, the wtd shows power-law behavior with similar
exponents, regardless of the threshold.
When the threshold $r_c$ is large, the total number of data points 
for the wtd decrease. Therefore, the wtd fluctuates much and
the uncertainty in the exponent $\beta_2$  increases.

\begin{figure}[h!]
\includegraphics[width=7cm,angle=0,clip]{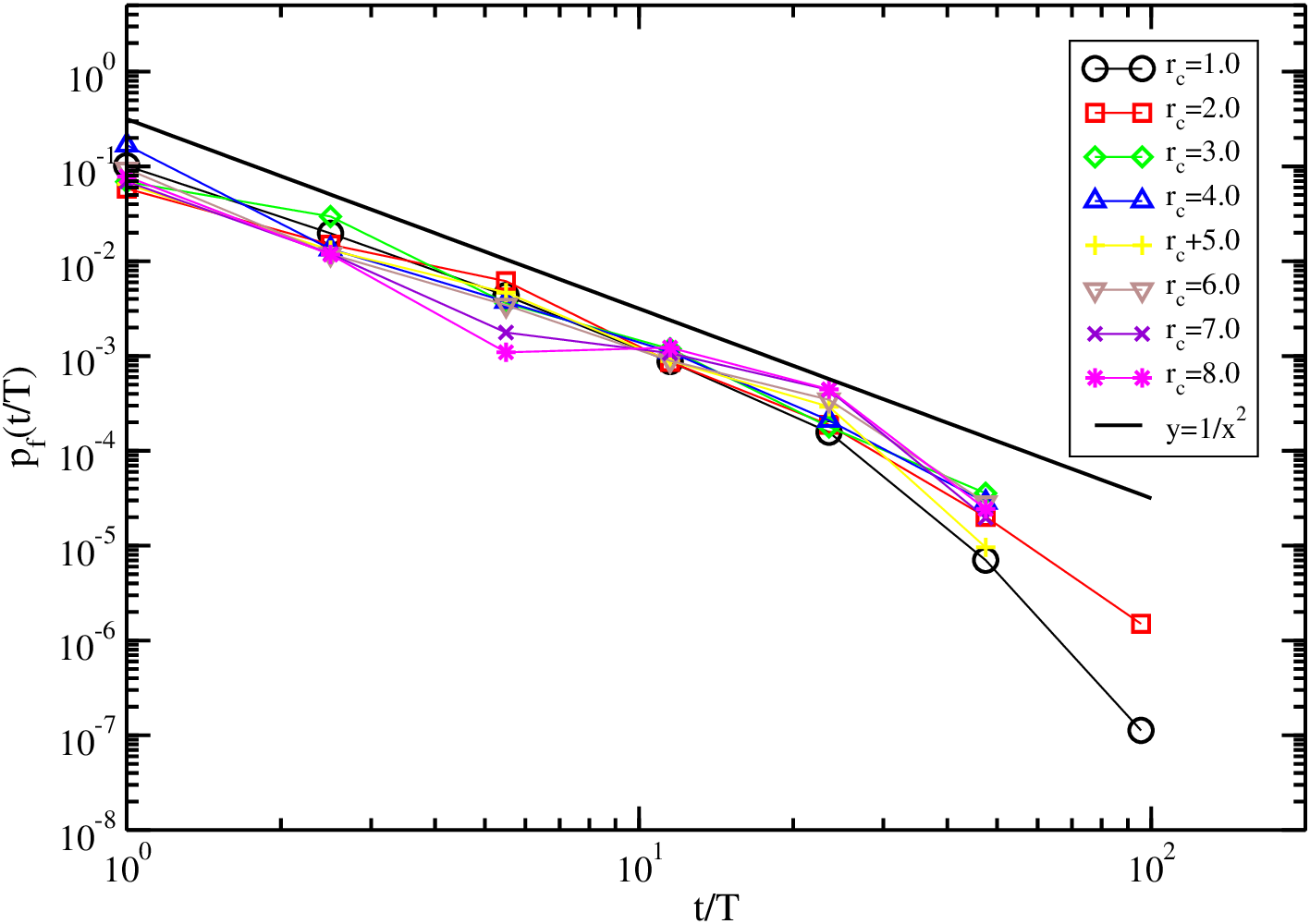}
\caption[0]{(Color online.)
Wtd of the absolute return 
scaled by the average waiting time $\langle T \rangle$ with $\Delta t=1$~min.
}
\label{fig3}
\end{figure}

The power-law exponents of the wtd are measured by least-squares 
fits. In Table~1 we present the exponents $\beta_2$ for $t>t_c$.
To measure the exponents, we scaled the
wtd by the average waiting time $ \langle T \rangle =\int_1^{\infty} t p_f (t) dt$. 
We present the wtd scaled by the average
waiting time in Fig.~\ref{fig3} for $\Delta t=1$~min.
The scaled wtd shows clear power-law behavior. 
For a given return time, the exponents $\beta_2$ are nearly equal 
within the error bars, regardless of the threshold $r_c$.
We also observe that the exponents $\beta_2$  decrease when the
return time $\Delta t$ increases.
We obtained the averaged exponents $\beta_2$ for the wtd as
$\beta_2 =2.0$ for $\Delta =1$~min, 1.58 for $\Delta t=60$~min,
and 1.42 for $\Delta t=600$~min.
It is very difficult to identify the origins of the scaling behavior
for the wtd. The correlation of the return is one reason of the
scaling behavior as shown in Fig.~\ref{fig1}, 
because the shuffled data set destroys the correlation of the time
series. The power-law behavior of the probability density function
for the absolute return is another reason for the scaling behavior
of the wtd. 
These power-law behaviors may be due to herding behavior of the
stock traders and the nonlinear dynamics of the stock market.

\begin{table}
\caption{Critical exponents $\beta_2$ for the wtd in the large-$t$ regime.
}
\begin{tabular}{c|c|c|c} \hline
$r_c$ & \multicolumn{3}{c}{ $\beta_2$ }    \\ \hline
      &  $\Delta t=1$~min &  $\Delta t=60$~min &
      $\Delta t=600$~min \\ \hline
1.0 & 2.04(7) &     & 1.40(5) \\ \hline
2.0 & 2.0(2)  &   1.58(7) & 1.52(7) \\ \hline
3.0 & 2.0(1)  &   1.6(1)  & 1.36(5) \\ \hline
4.0 & 2.1(1)  &  1.58(7) &         \\ \hline
5.0 & 2.1(3)  &  1.46(5) &         \\ \hline
6.0 & 2.0(1)  &          &         \\ \hline
7.0 & 1.9(2)  &          &         \\ \hline
\end{tabular}
\label{tab1}
\end{table}

\section{Conclusions}
We have considered the probability density function of the absolute return
and the waiting-time distribution (wtd) with a cut-off threshold. We 
observed that the probability density function of the absolute return
has a power-law behavior. The exponents $\alpha$ decrease
when the return time $\Delta t$ increases. 
We defined the waiting time of the absolute return by the threshold
$r_c$. The wtd also shows power-law behavior.
When the return time $\Delta t$ is less than one day, we observe
two scaling regimes, separated by a crossover time around $t_c \approx 200$~min.

\begin{acknowledgments}
This work was supported by KOSEF(R05-2003-000-10520-0). 
\end{acknowledgments}

\newcommand{\jpa}{J. Phys. A}
\newcommand{\jkps}{J. Kor. Phys. Soc.}

\end{document}